# Coherent Multidimensional Spectroscopy at the Few-Cycle Limit: Accounting for the Pump Spectrum

Jacob S. Higgins, Gerrit N. Christenson, James D. Gaynor

Department of Chemistry, Northwestern University, Evanston, IL

## Abstract

Coherent multidimensional spectroscopy will benefit greatly from near single-cycle pulses through the spectral breadth of states that can be accessed and the high frequency oscillations that can be resolved. However, the necessary cost of working with such pulses is the spectral structure arising from efficient self-phase modulation, which imparts into the signal and complicates interpretation of the molecular response. In this paper, we develop an *in-operando* approach to characterize and remove pump spectral features from two-dimensional spectroscopic signals. The pump spectrum is retrieved via an interferometric integrated autocorrelation measurement that runs concurrently with the experiment and thereby inherits the same time and frequency domain properties. A deconvolution procedure then suppresses noise and extracts the third-order response from the total signal. The analysis enables correction of spectral intensities and unambiguous assignment of spectral features to the molecular response. This work pushes multidimensional spectroscopy toward the impulsive limit and enables integration with other nonlinear spectroscopies such as attosecond spectroscopy – all without compromising few-cycle pulse properties.

## Introduction

Attosecond science is forging a new nonlinear optics toolkit to perform ultrafast time-resolved spectroscopies in chemistry, physics, and materials science.[1-7] In particular, tabletop high harmonic generation for attosecond pulse creation has been enabled by the routine generation of few-cycle laser pulses with multi-millijoule energies using hollow core fiber (HCF) spectral broadening schemes,[8-10] which are now commercially available. Few-cycle pulses are generated via self-phase modulation to create supercontinua spanning wavelengths from 450 nm to 1000 nm with sub-5 femtosecond (fs) Fourier-limited pulse durations.[11,12] These pulses are the engine of modern attosecond science, pushing the limits of experimental time resolution to directly monitor pure electronic motion on its natural timescale.

Their impressive bandwidth and few-femtosecond durations make few-cycle pulses ideal prospects for coherent multidimensional spectroscopy, where the time-frequency trade-off is circumvented to maintain instrument-limited temporal and spectral resolution of a material's nonlinear response. Coherent two-dimensional spectroscopy is a third order nonlinear technique with a powerful advantage of unveiling correlated energy gap fluctuations and quantum coherences from the molecular response function.[13-16] This method has been extensively used to study excited electronic dynamics in complex systems, including nanocrystals,[17-19] photosynthetic proteins,[20-22] semiconductors,[23,24] and photochemical systems.[25-27] Combining the tools of attosecond science with multidimensional spectroscopy will offer unprecedented windows into (de)localized and correlated electronic dynamics, unique many-body interaction signatures,[28,29] and quantum coherent phenomena. Improved spectral bandwidth could completely cover inhomogeneously broadened electronic excitation lineshapes to reveal competing excited state pathways that are not observable with narrowband excitation. Lineshape analyses, which can reveal complex system-bath correlations[30,31] and nonadiabatic dynamics[32,33] (e.g., at conical intersections) through nonequilibrium interactions between sub-ensembles, are also made more robust through resolution of the entire ensemble.

Despite their advantages, few-cycle pulses necessarily inherit spectral intensity modulations from self-phase modulation.[34] In multidimensional spectroscopy, the emitted signal field is the convolution of the third order molecular response and three incident electric fields. Major progress will be realized through procedures that account for their modulated spectral profile to meaningfully and accurately interpret multidimensional spectra. One potential remedy is amplitude pulse shaping for spectral smoothing. Although pulse shaping of similar broadband pulses has been shown for compression,[35-37] commercially available pulse shapers typically have limited shaping bandwidth, incompressible dispersion contributions, and limited input pulse energies with low shaping efficiencies. Another remedy, which is explored in this paper, is *in-operando* interferometric characterization of the few-cycle excitation pulse coupled with deconvolution algorithms to remove pump-induced spectral artefacts (**Figure 1**).

Here we develop such an approach to separate the molecular response from structured few-cycle spectra and enable the full capability of broadband, few-cycle two-dimensional spectroscopy. We demonstrate the method using two-dimensional electronic-vibrational (2D EV) spectroscopy.[32] The procedure measures an interferometric autocorrelation of the few-cycle pump and its transmission through the sample simultaneously with the 2D measurement to compare the visible spectral features and the nonlinear response under identical Nyquist conditions. We construct a Wiener filter that deconvolves the intensity and structure of the molecular response. Pump autocorrelation is measured before and after sample transmission to reconstruct the linear spectrum of the excitation band. We demonstrate the utility of this method with 2D EV spectra of two nanoparticle systems with systematically shifted absorption spectra. Pump spectral features are accounted for and largely deconvolved; the spectral intensity is corrected and purely reflects the oscillator strengths of the states being accessed. Our work provides a pathway for spectroscopists to interpret featured 2D spectra without sacrificing pulse power, bandwidth, temporal duration, or general quality of few-cycle laser pulses.

## Results

An experimental setup overview is shown in **Figure 2** (also described in *Methods*). Briefly, few-cycle pulses are generated via a HCF compressor, spanning 500-1000 nm with relatively flat phase and inherent spectral modulation due to efficient self-phase modulation. The entire pulse profile is broadened and temporally compressed to the few-cycle limit (4-5 fs). A high-precision Mach-Zehnder (MZ) interferometer produces a collinear pulse pair with time delay $\tau_1$. A third mid-infrared probe pulse follows the waiting time, $\tau_2$. The three light matter interactions generate a third-order polarization during the detection time, $\tau_3$, represented as[13]

$$P^{(3)}(\tau_1, \tau_2, \tau_3) \propto \int_0^\infty d\tau_3 \int_0^\infty d\tau_2 \int_0^\infty d\tau_1 \, R^{(3)}(\tau_1, \tau_2, \tau_3) E_3(t-\tau_3) E_2(t-\tau_3-\tau_2) E_1(t-\tau_3-\tau_2-\tau_1). \qquad [1]$$

The third-order signal is emitted collinearly and detected with the probe beam. The above equation represents a convolution of the molecular response, $R^{(3)}$, with the three fields. When plotted in the $(\omega_1, \omega_3)$ frequency domains, as in 2D spectroscopy, the pump spectral features impart into the spectrum along $\omega_1$, scaling with pump intensity (**Figure 1**).

**Integrated interferometric autocorrelation reproduces visible pump spectrum.** The MZ interferometer[38] has two usable outputs that are exactly $\pi$-radians phase shifted (see *Methods*). The "bright arm" output is the experimental pump beam, while the "dark arm" is used to measure the pump spectrum for spectral deconvolution (**Figure 2A**). The latter is achieved through an integrated interferometric autocorrelation measurement (i.e., of $g^{(1)}(\tau_1)$) collected simultaneously with 2D data. The intensity of the interfering pump pulses is measured as a function of $\tau_1$ and Fourier transformed to $\omega_1$ under identical conditions and data processing as the 2D data to retrieve the pump power spectrum. A second autocorrelation measurement is taken after sample transmission of the bright arm, denoted as the "sample arm," which is used to generate an *in-situ* linear absorption spectrum similarly to Ryu *et al*.[18]

The transient response of Prussian Blue analog (CoFe) nanoparticles[39] at $\tau_2$=1 ps is shown in **Figure 2B**. The negative signal at 2092 cm$^{-1}$ is an excited state absorption (ESA) representing a CN stretch probed on the electronic excited state. The positive signal at 2126 cm$^{-1}$ is a ground state bleach (GSB) reporting on the loss of the ground state vibrational absorption in the presence of the pump pulse. Three 2D EV replicates of CoFe were taken at five $\tau_2$ points.

The dark arm is plotted alongside the peak of the GSB feature as the $\tau_1$ delay is scanned (**Figure 2B**). The two scans evolve with a similar frequency and envelope, indicating the clear convolution between the pump spectrum and the nonlinear response. In **Figure 2C**, the dark arm and sample arm are plotted together. The two waveforms are $\pi$-phase shifted, as required by the MZ interferometer, but otherwise have nearly identical phase and envelope evolution. The difference between their spectral profiles is the electronic absorption of the CoFe sample in the sample arm.

Through convolution with the pump, the 2D EV spectra have spectral features resulting from broadening in the HCF. **Figure 3A** shows a spectrum of CoFe nanoparticles taken at waiting time $\tau_2$=1 ps, plotted here as a single replicate to emphasize experimental signal-to-noise (see *Methods*). The excitation lineshape spans over 6,000 cm$^{-1}$. This is over twice the range of high frequency vibrational quanta in molecules, demonstrating the ability of few-cycle pulses to photoexcite broad excitation bands. However, many features are present along the excitation axis due to the pump structure, complicating spectral interpretation and necessitating deconvolution.

Accurate pump spectrum deconvolution relies critically on using the *in-operando* pump autocorrelation because it is collected under identical time steps as the 2D data. The pump spectrum is retrieved by Fourier transforming the dark arm autocorrelation measurement (**Figure 3B**). The spectrum is nearly identical to a separate grating-resolved measurement (**Figure 3B**), demonstrating the accuracy of the pump reconstruction protocol. Slight deviations in the structure and peak resolution arise from imperfect calibration of the wavelength-based spectrometer. Deconvolution with the wavelength-based spectrometer would improperly retain structure at several frequency bins and contribute additional spectral artefacts, preventing accurate assessment of the desired signals.

The dark arm measurement accurately accounts for pump spectral features in the 2D spectrum. **Figure 3C** shows the dark arm plotted with the same GSB slice Fourier transformed to $\omega_1$. The structure in the 2D spectrum is reproduced exactly in the dark arm spectrum due to their concurrent measurements in $\tau_1$. The two spectra vary in intensity across $\omega_1$ because the dark arm measures only the pump spectral intensity while the nonlinear signal measures the pump intensity convolved with the molecular response function.[13,14] Thus, the intensity ratio between the dark arm and GSB slice report on the oscillator strength of the excitation band. Plotted in black is the CoFe linear absorption spectrum. The transition dipole strength is stronger at higher frequencies. Consequently, the blue edge of the excitation band at 16,000 cm$^{-1}$ is stronger in the 2D EV spectrum despite having similar pump magnitude across the entire bandwidth.

**Wiener deconvolution retrieves the pure molecular response.** Naively, one should be able to deconvolve the molecular response by simply dividing the pump spectrum from the 2D data at each $\omega_3$ point, per the convolution theorem.[13] However, the presence of noise and modulations in the pump spectrum cause the dark arm amplitude to approach or cross zero at certain frequencies. Hence, division greatly amplifies noise in low signal-to-noise regions in both the dark arm and 2D spectrum. This common deconvolution problem is solved through the Wiener deconvolution procedure, developed to minimize the mean-squared-error of the deconvolved signal with respect to the true solution while attenuating the noise.[40,41] In effect, Weiner deconvolution divides the pump from the signal in high signal-to-noise frequency bins and suppresses the signal in low signal-to-noise bins where noise would otherwise be amplified.

The Wiener filter is constructed using the following equation:

$$WF(\omega_1, \omega_3) = \left[\frac{1}{pump(\omega_1)}\right] \frac{1}{1+\alpha[NSR(\omega_1,\omega_3)/pump^2(\omega_1)]} \qquad [2]$$

The term $pump(\omega_1)$ is the spectral intensity of the pump beam. The first bracketed term is simple pump division. The second term accounts for noise; the term $NSR(\omega_1, \omega_3)$ is the noise-to-signal ratio of the spectra. The bracketed term in the denominator pushes the signal toward zero in bins with small pump amplitude or signal intensity. The value $\alpha$ is a tuning parameter that controls how strongly the noise is suppressed (see *Methods*).

An overview of the Wiener deconvolution algorithm is diagrammed in **Figure 4A** and detailed in *Methods*. An $\alpha$ value for the entire dataset is chosen to balance deconvolution with noise amplification (**Figure 4B**). Appropriate 'signal' and 'noise' regions are then chosen in the $\omega_1$ dimension. For a given replicate $i$, the noise profile is extrapolated across $\omega_1$. The noise-to-signal ratio is calculated for every $\omega_1$ and $\omega_3$ point in the 2D spectrum and averaged across $\tau_2$. Because the Wiener filter suppresses the signal in the presence of noise, it only depends on $\omega_1$ and $\omega_3$; it is $\tau_2$-independent to prevent artificially damping kinetic signals as their amplitudes change through the waiting time.

The spectral correcting capability of the Wiener filter is demonstrated for a single replicate of the CoFe data – a noisiest-case scenario due to no signal averaging. A representative Wiener filter and deconvolved 2D spectrum is plotted in **Figure 4C**. The Wiener filter intensity (left) shows the structure inherent to the pump and approaches zero in low signal regions. The filtered 2D spectrum (right) therefore has reduced pump structure and lower noise than the unfiltered spectrum. The filter corrects for the transition dipole intensity that is stronger at the blue edge of the electronic excitation for CoFe, as observed in the linear absorption spectrum (**Figure 3C**). The filter and resulting 2D spectral strength are consistent with the linear absorption spectrum: strongest from 14,000 to 16,000 cm$^{-1}$ and approaching zero above 16,000 cm$^{-1}$ where there is low pump spectral intensity. The reduction of noise and correction for spectral structure demonstrate this approaches' utility for deconvolution, enabling coherent multidimensional spectroscopy with attosecond and few-cycle laser sources.

**Amplitude-corrected nonlinear response and hidden pump-probe correlations.** The weighted average that follows the deconvolution algorithm maximally reduces the pump influence on the 2D spectrum. **Figure 5A** shows a fully deconvolved 2D spectrum of CoFe where the >6,000 $cm^{-1}$ excitation bandwidth is not compromised. The noise is significantly reduced relative to a single replicate with minimal averaging (**Figure 4C**) and is almost completely absent in regions with no pump intensity or vibrational resonance along the probe axis ($\omega_3$ <2080 $cm^{-1}$ and $\omega_3$ >2160 $cm^{-1}$). The excitation axis is smoothed without frequency-binning or boxcar averaging, and the pump features are mitigated relative to the unfiltered spectrum. Crucially, the remaining spectral features that are not due to the molecular response are properly and unambiguously accounted for in the dark arm spectrum. The spectral amplitude across the $\omega_1$ axis follows the molecular response consistent with the linear electronic absorption. Importantly, the population evolution dynamics are not affected by the deconvolution procedure, shown through a match between the $\omega_1$-integrated 2D spectrum and the transient IR spectrum (**Figure 5B**), consistent with the projection slice theorem.

The sample arm interferometric autocorrelation is used to construct a Fourier transformed visible linear absorption spectrum. This spectrum has the same deconvolution procedure applied with the same $\alpha$. Comparing the sample arm to the $\omega_1$-dependent lineshape of 2D EV spectral slices can reveal time-dependent correlations between the electronic pump and vibrational probe transitions, see *Discussion*. The trends in the filtered spectrum match the linear transmission spectrum (**Figure 5C**), showing higher absorption at the blue edge. An important self-consistency check is built-in to this approach: The spectral holes at 12,000 and 13,500 $cm^{-1}$ can also be seen in the deconvolved 2D spectrum and dark arm, which can therefore be ascribed to low signal-to-noise regions in the Wiener filter rather than the molecular response.

To demonstrate that the deconvolution procedure isolates only the molecular response, 2D spectra were also taken for Prussian Blue (FeFe) nanoparticles, which feature a red-shifted electronic absorption compared to CoFe. **Figure 6** shows a deconvolved FeFe spectrum collected using an identical pump spectrum as the CoFe sample. For the FeFe spectrum, the deconvolved spectral intensity is flatter across $\omega_1$ than CoFe (**Figure 6B**), demonstrating that the transition dipole strength is more evenly distributed across the excitation band, as expected by the flatter transmissivity of the *in-situ* sample arm linear spectrum of FeFe (**Figure 6C**). The consistency between these spectral profiles demonstrates the accuracy in deconvolving the pump spectrum from the molecular response.

## Discussion

Few-cycle pulses offer unprecedented bandwidth and temporal resolution for multidimensional and attosecond spectroscopy. As the commercial availability of few-cycle pulse sources expands their user-base and application space, it is imperative that technical advances are simultaneously developed to account for their inherent spectral structure. Although few-cycle pulse spectral structure is less important than the electric field's peak magnitude in strong field attosecond physics research, the spectral profile is exceedingly important where resonant electronic transitions are excited in ultrafast nonlinear spectroscopies. This work overcomes this challenge through *in-operando* measurement of the pump spectral structure and subsequent extraction of the pure molecular response, allowing unambiguous and accurate interpretation of multidimensional spectra. We demonstrate the procedure with 2D EV spectroscopy, though it is applicable to any coherent multidimensional spectroscopy with structured pump pulses.

The high bandwidth (>6,000 cm$^{-1}$) pulses generated here bring ultrafast multidimensional spectroscopy closer to the impulsive limit, where the pulses can be broader than the interrogated electronic absorption lineshapes and the third-order polarization approaches the exact molecular response (**Equation 1**). In the time-domain, 4-5 fs pulse durations enable resolution of high frequency coherences during $\tau_2$, and in-phase excitation that occurs at least twice as fast as any vibrational motion. In the frequency-domain, excitation spectra surpass the linewidth of most electronic excitations in condensed phase molecules and cover large swaths of the excitation bands in materials (**Figure 3**). For comparison, two-dimensional infrared spectroscopy benefits from commercially available femtosecond IR pulses that broadly span infrared-active stretching bands.[13,42-45] This capability allows simultaneous excitation of all sub-ensembles within an absorption so that dynamic correlations in energy gap fluctuations and coherent energy transfer can be robustly extracted. Similarly, the developments in this paper are necessary for few-femtosecond and attosecond multidimensional spectroscopies to characterize electron-electron correlations and quantum coherences in complex environments.

The *in-operando* measured pump spectrum provides the spectral information needed to unambiguously identify and deconvolve pump-induced spectral artefacts through Wiener deconvolution. Though some spectral features remain in the deconvolved spectra, amplitudes of the molecular response are fully corrected, and pump spectrum artefacts are unambiguously identified to prevent misinterpretation. Pump features are strongly mitigated relative to the pre-filtered data such that unique molecular signatures can be discerned through differences in the spectral artifacts in the dark arm. The signal-to-noise of the filtered spectra are greatly improved (**Figures 5 & 6**) due to the Wiener deconvolution algorithm's judicious suppression of noise in frequency bins that lack appropriate signal strength.

The ability to recover broadband excitation profiles reveals hidden correlations between excited and detected degrees of freedom that are inaccessible to narrowband pulses in conventional ultrafast experiments. Importantly, perfect correlation between the pumped transitions (e.g., electronic) with the probed transitions (e.g., vibrations) should not be assumed *a priori.* This is only borne out in quantitative differences between the structure of the 2D excitation band and the *in-situ* transmission spectrum measured through the sample arm. The above 2D EV spectra are instructive examples: For a given vibration probed along the detection axis, the ensemble of electronic excitations might differently overlap with, or modify, the oscillator strength of the vibrational coordinates being probed. For CoFe, matches between these spectra (**Figure 5D**) indicate near perfect correlation. Intensity mismatches due to imperfect correlation between electronic and vibrational ensembles will reveal key insight regarding how ultrafast electronic rearrangement modifies the vibrational response – information that sensitively reports on excited state dynamics. This example is generalizable, especially for spectroscopies where the pump and probe spectra differ in photon energy (i.e., "off-diagonal" spectroscopies).

This method establishes a framework for complete reconstruction of the nonlinear response. The initial wavepacket preparation and coherent excited state dynamics are affected by the phase of the excitation pulse,[46,47] which can be directly incorporated with more in-depth dark arm characterizations. Phase retrieval protocols such as TG-FROG[48] can capture the pump chirp and allow additional reconstruction of time axes.[49] Deconvolution models based in deep learning algorithms[50] or bottom-up reconstruction of the constituent Feynman pathways[51,52] can concurrently identify the underlying molecular features and quantify time-dependent correlations. The general methodology established here integrates coherent multidimensional spectroscopy with state-of-the-art tools from attosecond science and circumvents technical limitations of pulse shapers, which can ultimately be extended to probe pulses generated by high harmonic generation. Deconvolution permits broad bandwidth, few-cycle two-dimensional spectroscopy without compromising pulse power or spectral resolution, enabling the full capabilities of multidimensional spectroscopic analysis.

## Figures

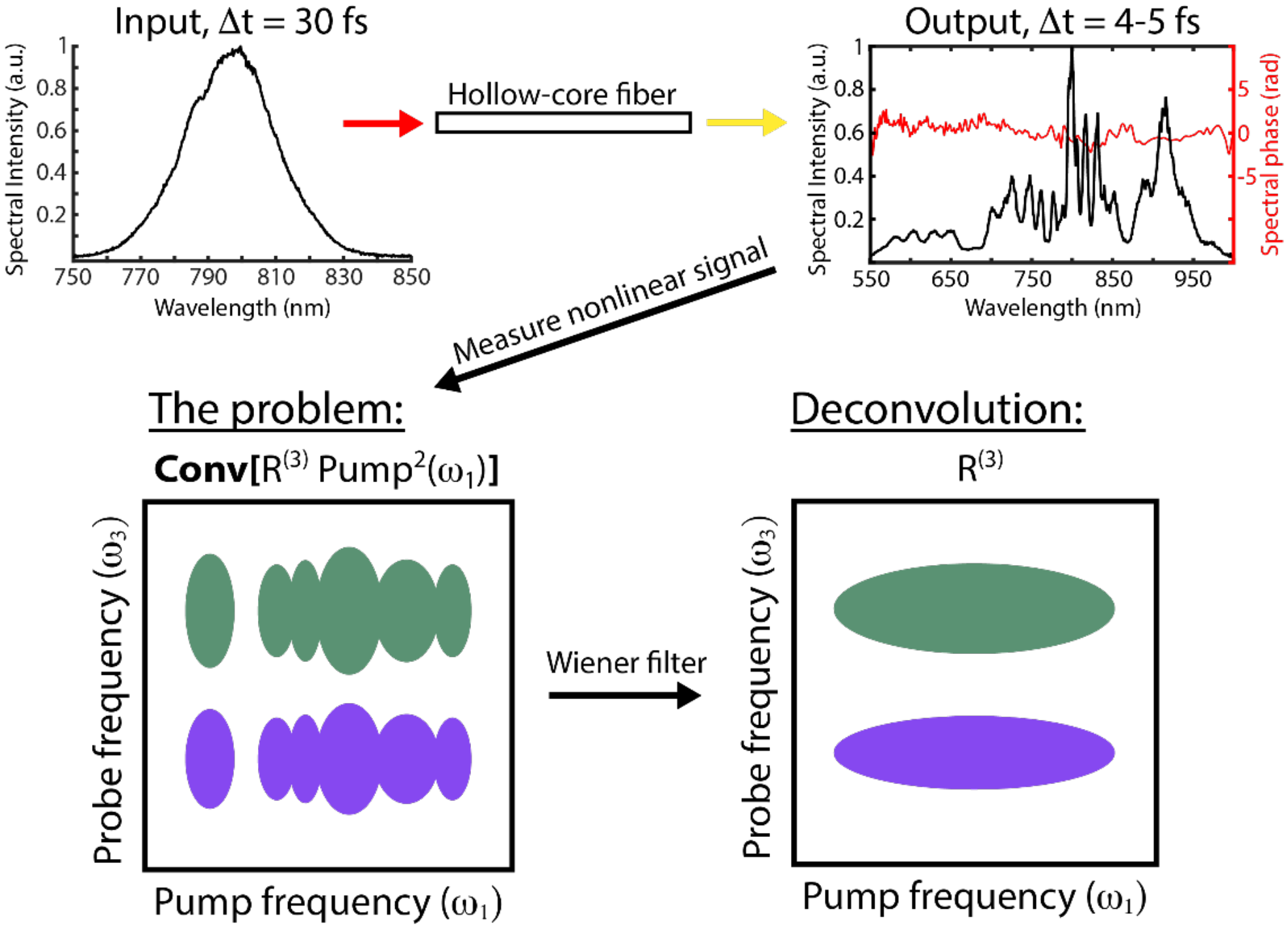

**Figure 1. Overview of the problem.** A hollow-core fiber is used to spectrally broaden ultrafast pulses to near single-cycle temporal durations, but with added spectral structure. These features are imparted into the nonlinear response and lineshape of a two-dimensional spectroscopy measurement, necessitating a deconvolution procedure to remove the spectral structure and isolate the pure molecular response, $R^{(3)}$.

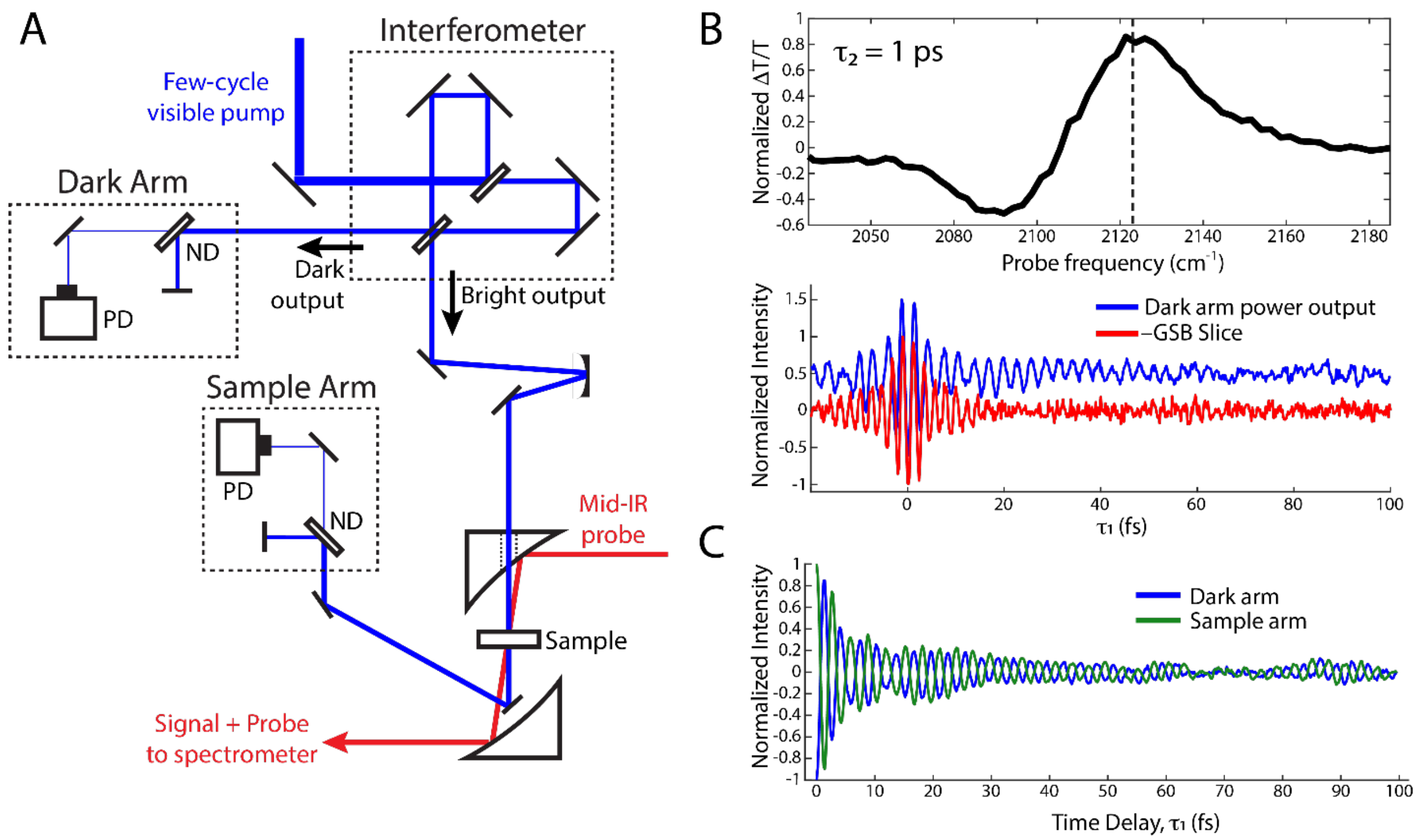

**Figure 2. Data collection scheme.** A) A Mach-Zehnder interferometer is used to split the broadband, few-cycle pump pulse into two pulses with a controllable time delay, $\tau_1$. The dark arm output is used to measure the pump spectral intensity via a Fourier transformed integrated autocorrelation measurement. The output of the bright arm is used for the 2D EV measurement. After transmission through the sample, the pump pulses are measured in an additional autocorrelation measurement. B) Transient infrared spectrum of CoFe nanoparticles taken at $\tau_2$=1 ps (top). This ground state bleach feature is modulated as a function of $\tau_1$ with a similar temporal profile as the dark arm measurement (bottom), offset for clarity. C) The dark arm and sample arm modulations as a function of $\tau_1$ show nearly identical temporal evolutions with phase offset $\pi$ and difference due to sample absorption.

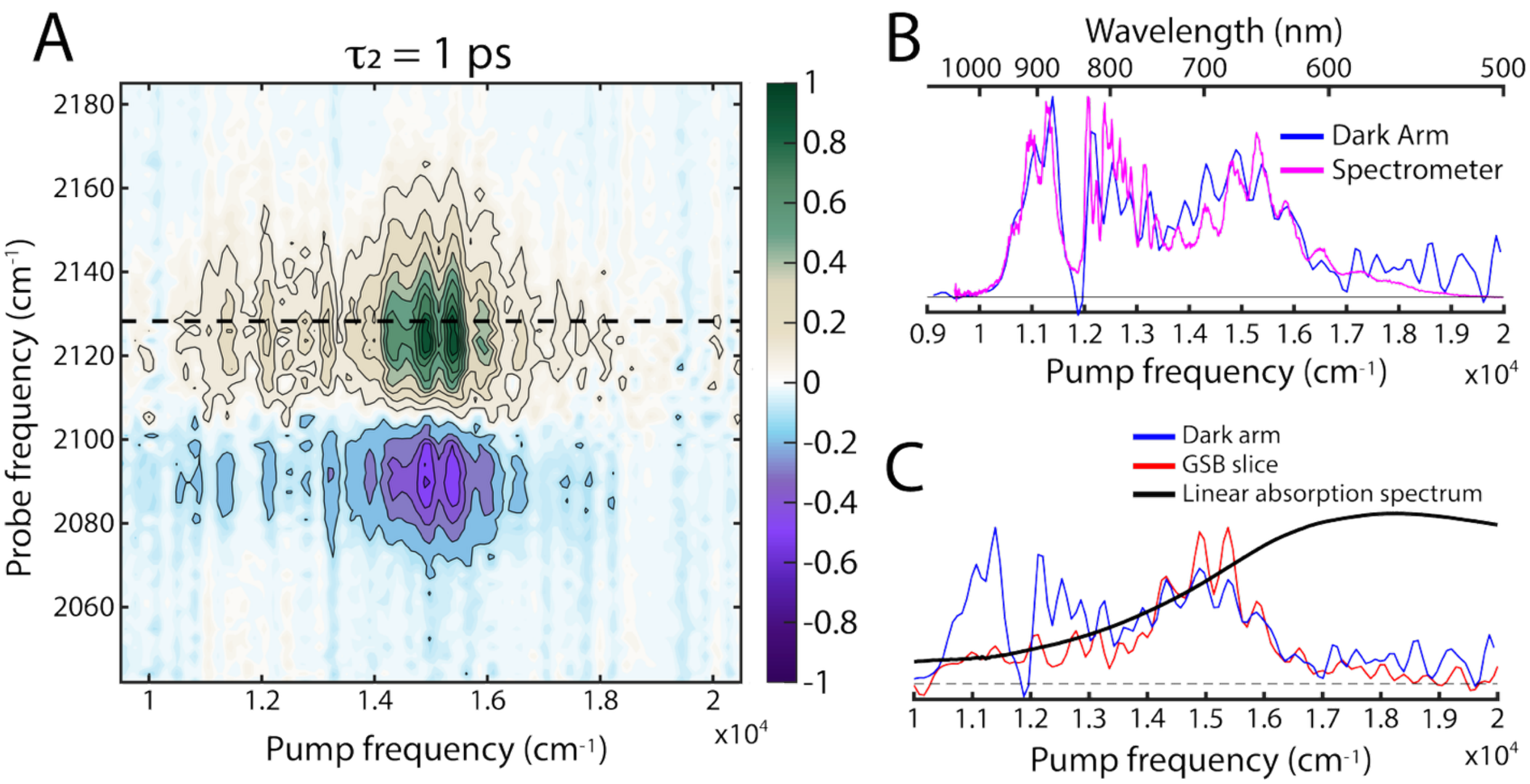

**Figure 3. Reproduction of pump spectrum through dark arm measurement.** A) A single replicate of a 2D EV spectrum of CoFe nanoparticles at waiting time delay $\tau_2$=1 ps. The spectral features across the pump frequency are largely due to the structure of the broadband pump pulse. B) The dark arm measurement taken alongside a grating spectrometer measurement of the pump show similar spectral features, demonstrating the accuracy of the pump reconstruction. C) The Fourier transformed dark arm measurement shares the same pump-dependent spectral features as the 2D spectrum, demonstrated here by plotting the dark arm with a slice of the 2D EV spectrum taken along $\omega_1$. The linear absorption spectrum is plotted to show the excitation transition dipole strength.

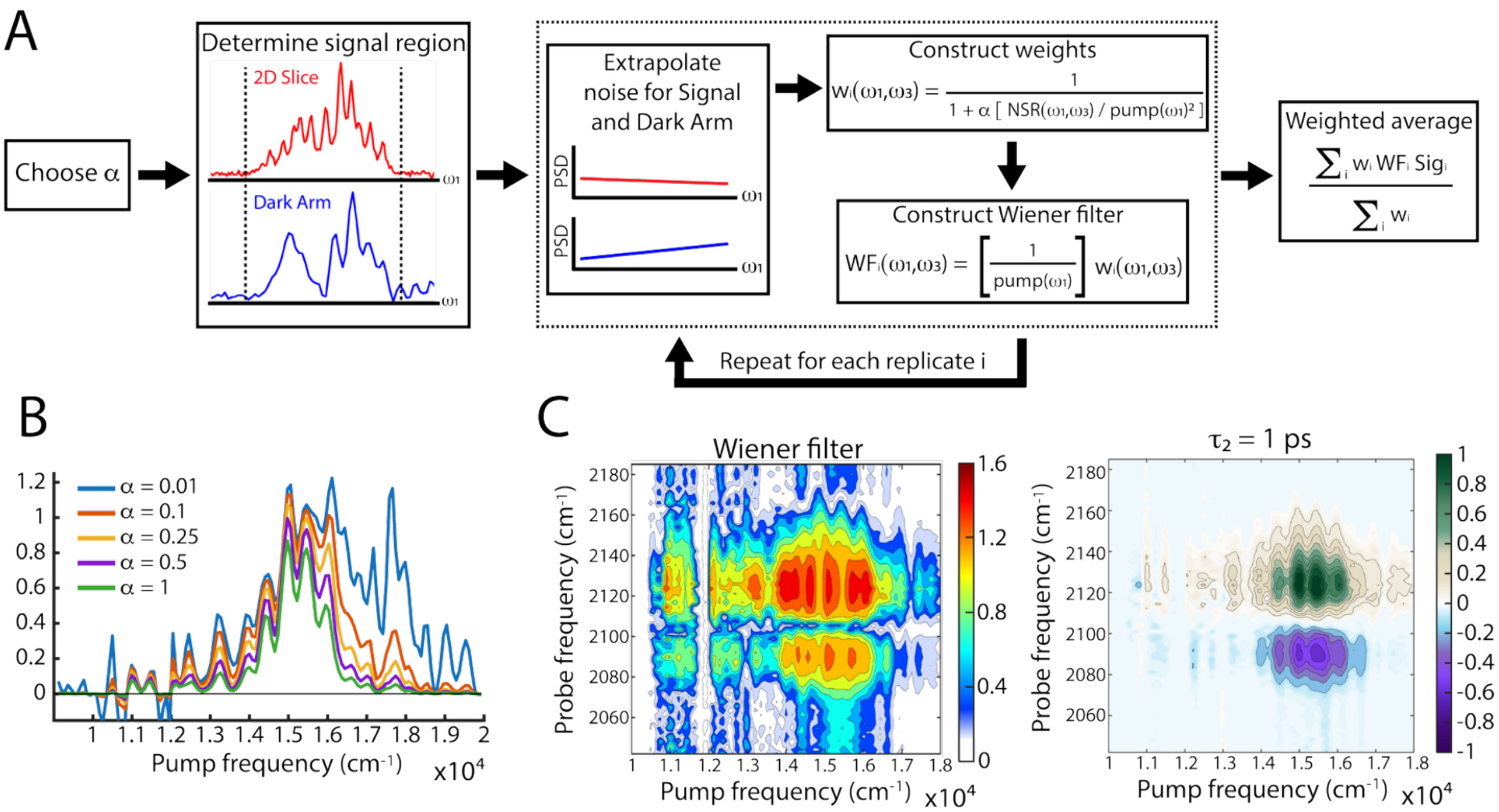

**Figure 4. Overview of deconvolution procedure.** A) Schematic of the Wiener deconvolution algorithm. After choosing the parameter $\alpha$, which determines how strongly the filter suppresses noise, and choosing the signal region of interest, the algorithm estimates the noise power spectral density via extrapolation through the signal region and generates the weights and Wiener filter function for each replicate $i$, or 2D data cube. A weighted average is then taken to generate the final spectrum. B) A deconvolved ground state bleach slice is plotted as a function of $\alpha$. Low $\alpha$ values amplify spectral noise, while high $\alpha$ values suppress the signal. C) Exemplary Wiener filter and deconvolved spectrum for a single replicate plotted at $\tau_2$=1 ps for $\alpha$=0.25. The filter and weight function depend on $\omega_1$ and $\omega_3$ but not $\tau_2$.

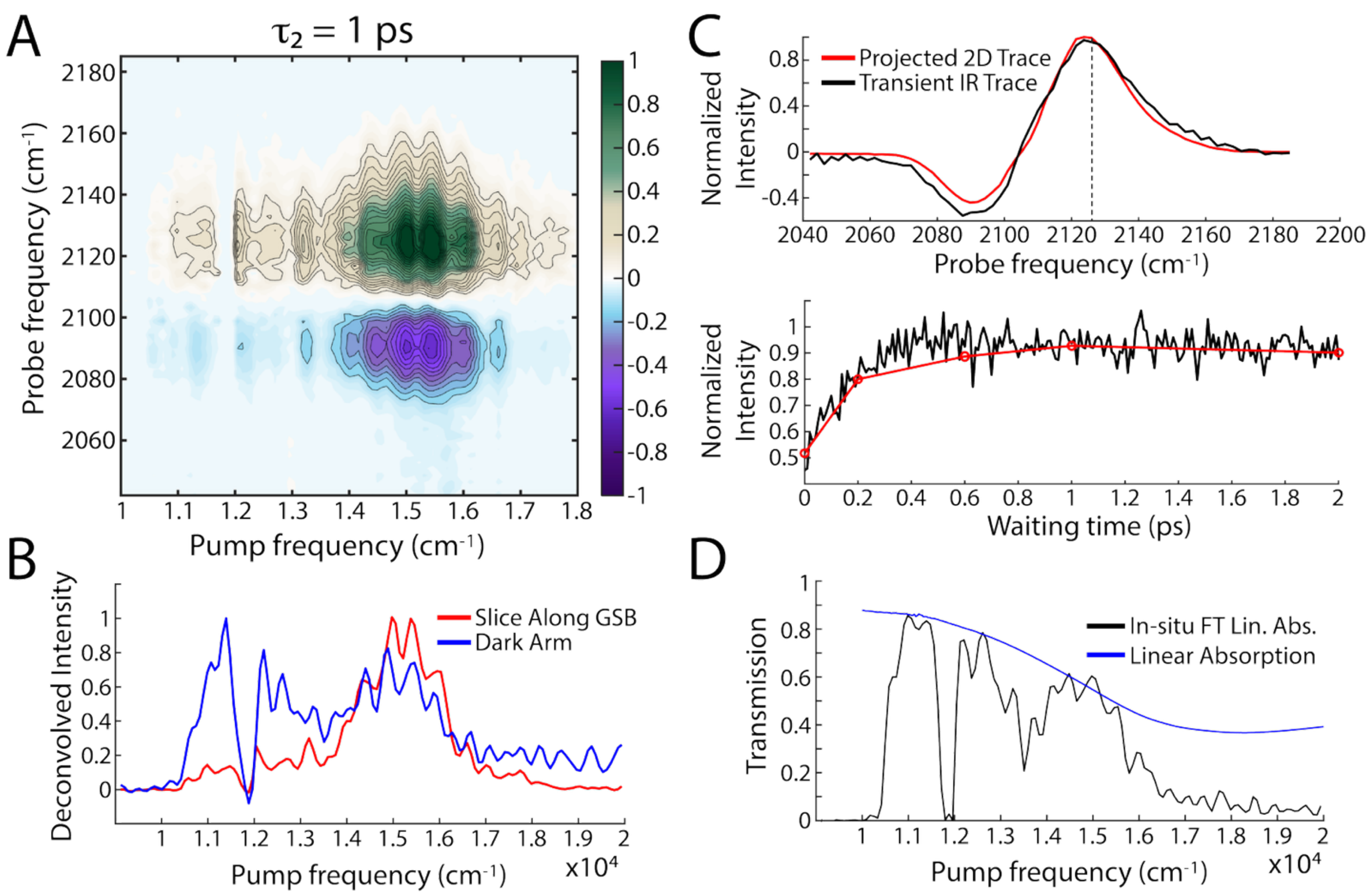

**Figure 5. Fully deconvolved 2D spectrum of CoFe.** A) The averaged, deconvolved 2D EV spectrum with α=0.25 plotted at $\tau_2$=1 ps has mitigated pump spectral features and spectral intensity that is purely due to the molecular response. B) The same ground state bleach slice as Figure 3 is plotted alongside the dark arm to demonstrate corrections of spectral amplitude, which match the transition dipole strength of the CoFe electronic excitation band. C) The transient IR spectrum (top) is plotted alongside the 2D EV spectrum summed over the pump frequency axis (bottom) to demonstrate that deconvolution does not affect signal kinetics along $\tau_2$. The measured $\tau_2$ points of the projected 2D spectrum are $\tau_2$=0, 0.2, 0.6, 1, and 2 ps. D) The dark arm and sample arm measurements are used in a T/$T_0$ measurement using the Wiener deconvolution procedure, which largely matches the linear absorption spectrum, plotted as transmission, in overall intensity trend.

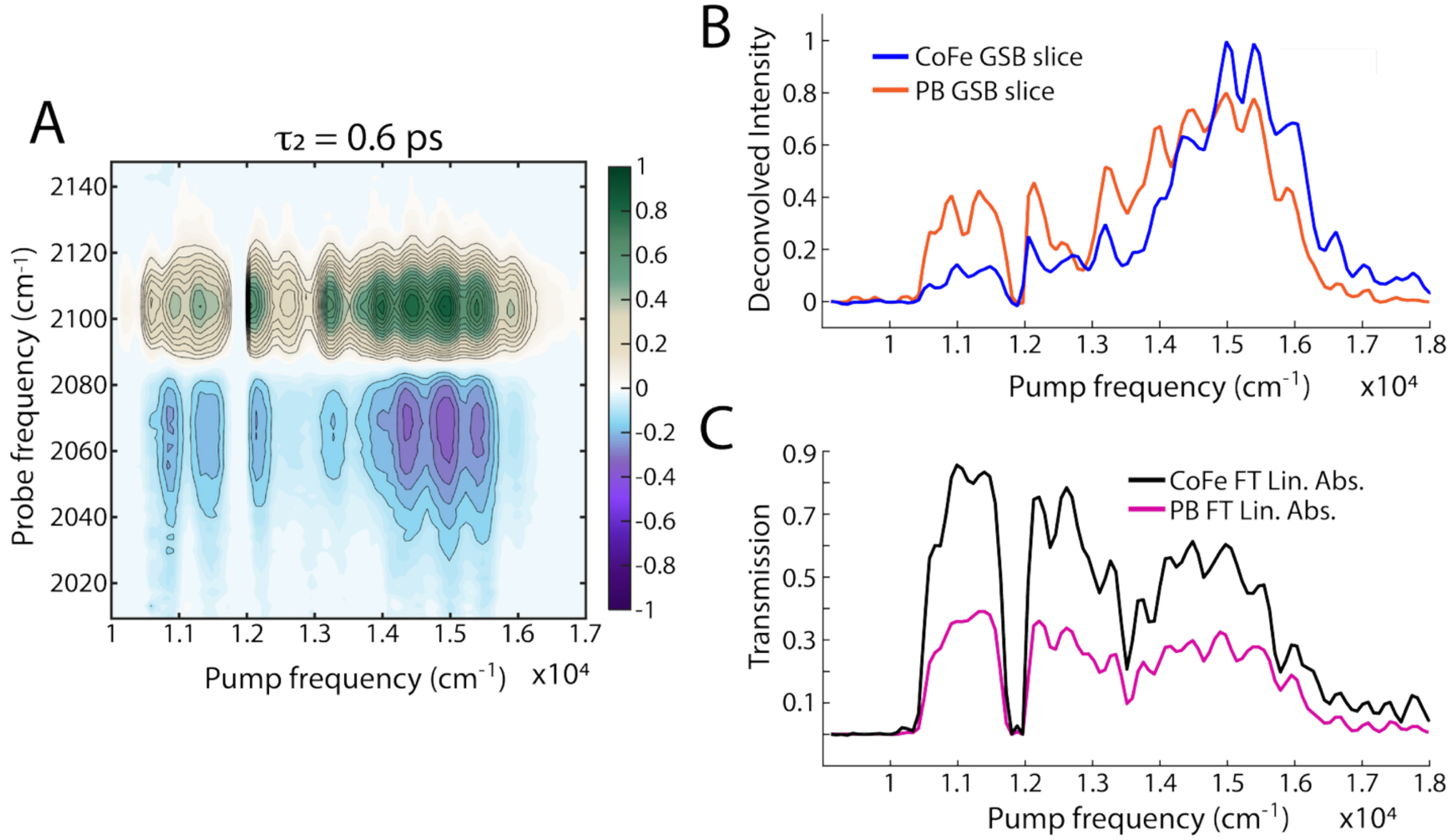

**Figure 6. Fully deconvolved 2D spectra isolate the molecular response absent the pump.** A) Averaged 2D EV spectrum of Prussian Blue nanoparticles with $\alpha$=0.15, plotted at $\tau_2$=600 fs. B) Deconvolved GSB slices of CoFe and Prussian Blue taken with the identical pump spectra. The different relative amplitudes between the two slices across $\omega_1$ are due to the molecular response. This is supported by C) the *in-situ* generated linear absorption spectra of CoFe and Prussian Blue. Prussian Blue has a relatively flatter transmissivity across the pump spectrum, while CoFe is relatively less transmissive (more absorptive) at the blue edge.

## Methods

**Ultrafast measurement.** The apparatus used to perform 2D EV spectroscopy is similar to the transient-IR apparatus described previously[39], with several developments described here. Briefly, a Ti:sapphire chirped pulse amplifier (Coherent Astrella, 7W, 1 kHz, ~33 fs) output is separated into two 3.5 mJ/pulse beamlines to generate the few-cycle pump pulse and the femtosecond mid-IR probe pulse, respectively. The few-cycle pump beam is generated by focusing ~3.0 mJ into a pressurized 2-m long stretched HCF (500 μm inner diameter, 1.9 bar Ne static) to produce spectrally broadened pulses of ~1.9-2.2 mJ / pulse (500-1000 nm bandwidth). The pulses are consequently compressed using 12 bounces on double-angle chirped mirror pairs (Ultrafast Innovations PC1332) and bulk compression to fine tune dispersion using AR coated fused silica wedges (Layertec) and higher order dispersion compensation[12] using z-cut ADP crystals (United Crystals). The few-cycle excitation pulses are characterized under identical conditions to the sample area using D-cycle (Sphere Ultrafast Photonics), which is used to optimize the compression to near FT-limited pulse durations with flattest spectral phase possible accounting for all optical components prior to the sample (including front side $CaF_2$ sample cell window and broadband specialized beamsplitters in the MZ interferometer).

The attosecond MZ interferometer follows a similar design from the attosecond science literature with active phase stabilization capabilities.[38] The few-cycle beam is routed into the MZ interferometer to produce collinear phase-locked pump pulse pairs with a controllable $\tau_1$ delay using a piezoelectric nanopositioner stage (Physik Instrumente, P612.2 XY). The “bright arm” of the interferometer has perfectly constructive interference between the two pump pulse pairs at $\tau_1$=0 due to the two beams having balanced reflections and transmissions in the interferometer. Conversely, the “dark arm” of the interferometer has perfectly destructive interference between the two pump pulse pairs at $\tau_1$=0 due to the two beams having unmatched reflections and transmissions. This yields the perfect π-phase shift between bright and dark arms that is exploited in this approach, following naturally as a consequence of using free-space interferometry in a MZ geometry.

The bright arm pump pulse pair is then routed to the sample area for spatial and temporal overlap with the mid-IR probe. The mid-IR probe pulse is generated with the other half of the chirped pulse amplifier output routed through an optical parametric amplifier (TOPAS-Prime, Light Conversion) and difference frequency generation module (NDFG, Light Conversion). The mid-IR probe is routed to the sample area to overlap with the pump pulses through a computer-controlled delay stage (Newport ILS250BPP) to control the $\tau_2$ delay time. We operate our 2D EV spectroscopy in the pump-probe geometry, where the absorptive two-dimensional signal is generated colinearly with the infrared probe which is spectrally dispersed onto mercury-cadmium-telluride pixel array.[39] The nonlinear signal emerges from measuring the difference of this spectrum in the presence and absence of the pump and dividing the probe transmission ($\Delta T/T$). In the data presented here, the pump and probe beams have orthogonal electric field polarizations.

**Autocorrelation measurements.** An integrated autocorrelation measurement is taken in two areas of the experiment: the dark arm output of the MZ interferometer and after passage through the sample. At each arm, the two pump beams are collinear and interfere as the stage is scanned in $\tau_1$. The broadband photodetector used (ThorLabs PDA20X2) has a high bandwidth to allow for single-shot measurements of the 1 kHz pulse train. This detector is meant for higher repetition frequency pulse systems and thus has a low pulse peak power damage threshold. Attenuating the two arms using reflective ND optics (total ND of 6) was necessary, producing ~pJ pulse pairs. The ND optics and photodetector have wavelength-dependent responses which must be corrected for in post-processing.

Self-phase modulation in the HCF produces broadband pulses which are phase shifted across the wavefront. Focusing the full beam can therefore cause destructive interference in certain spectral regions. The holes in the pump spectrum would be accentuated in both the autocorrelation measurements and in the nonlinear signal (see **Figure S1**). Before alignment into the MZ interferometer, the beam is irised by a factor of 2-3 to a power of ~1.25 μJ per pulse. Spectral holes are monitored through processing the dark arm spectrum and minimized.

**Signal processing and Wiener deconvolution procedure.** At $\tau_1$=0 ps, the dark arm should have minimal intensity, and the sample arm should have maximal intensity through interference in the MZ interferometer. These points are used to find the $\tau_1$=0 point in both data collection and post processing. The dark arm, sample arm, and 2D data are interpolated between time bins in $\tau_1$ to find the maximum sample arm signal. After interpolation, the same $\Delta\tau_1$ time steps are used as the raw data to maintain the same frequency axis as the original measurement.

Subsequent processing of two-dimensional data follows typical procedures. For each waiting time point, the infrared signal is modulated as a function of $\tau_1$. The long-time signal amplitude is subtracted, and the signal envelope is apodized using a hyperbolic tangent window prior to Fourier transformation. Because the autocorrelation measurements are identical for all $\tau_2$ traces, they are averaged together in the $\tau_1$ domain for each 2D data replicate collected. Signal processing of the autocorrelation measurements uses the same apodization window and zero padding length as the 2D processing. After Fourier transforming the autocorrelation traces, the spectra are divided by the frequency-dependent efficiency of the ND filters and photodiode to correct for spectral losses in the autocorrelation branch (see **Figure S2**). The correction correctly produces the spectral amplitudes of the pump, shown in **Figure 3B** and discussed in *Results*. Due to the relative inefficiency at the blue spectral edge both in the ND and photodiodes, the blue edge of the spectrum is amplified both in signal and noise. This necessitates the consideration of both the dark arm and 2D spectral noise profiles in the Wiener deconvolution procedure.

Wiener filters are generated and independently applied for each repeated 2D spectrum measurement. For replicate $i$, the noise is extrapolated through the spectra via a linear fit to the noise regions, which are selected above and below the signal regions (**Figure 4**). The extrapolated noise profiles are combined between the dark arm and nonlinear signal through quadrature addition:

$$noise_{total} = \sqrt{noise_{Dark}^2 + noise_{Sig}^2}. \qquad [3]$$

Both noise profiles are considered to account for differences in relative noise across frequencies. After extrapolating the noise for replicate $i$, the weight function $W_i(\omega_1, \omega_3)$, and the Wiener filter $WF_i(\omega_1, \omega_3)$ are calculated. After the filters are constructed and applied for all replicates, they are combined in a weighted average with $W_i$ serving as the weights for each replicate. The weighted average privileges 2D spectral bins with stronger signal-to-noise values across the replicate measurements, which further improves the signal strength.

The $\alpha$ parameter in the Wiener deconvolution procedure is chosen by looking at the response of the spectral holes where there is low dark arm intensity. **Figure 3C** shows the GSB slice described in *Results* but subject to Wiener deconvolution with different $\alpha$ parameters. High values at 1 or above suppress the noise and minimally reduces the spectral features of the dark arm. Low values at or below 0.1 remove much of the spectral structure but significantly amplify the noise. For example, at $\alpha = 0.01$, there are prevalent spikes near the hole region of the dark arm at 12,000 cm$^{-1}$, which emerge from noise-induced zero crossings of the dark arm spectrum. An $\alpha$ value between 0.1 and 0.5 typically satisfies an intermediate condition where pump spectral features are corrected without significantly amplifying noise. Importantly, the $\alpha$ value is chosen for all replicates $i$ so that the weight functions and filters are constructed consistently.

# Supplementary Information For:

## Coherent Multidimensional Spectroscopy at the Few-Cycle Limit: Accounting for the Pump Spectrum

Jacob S. Higgins, Gerrit N. Christenson, James D. Gaynor

Department of Chemistry, Northwestern University, Evanston, IL

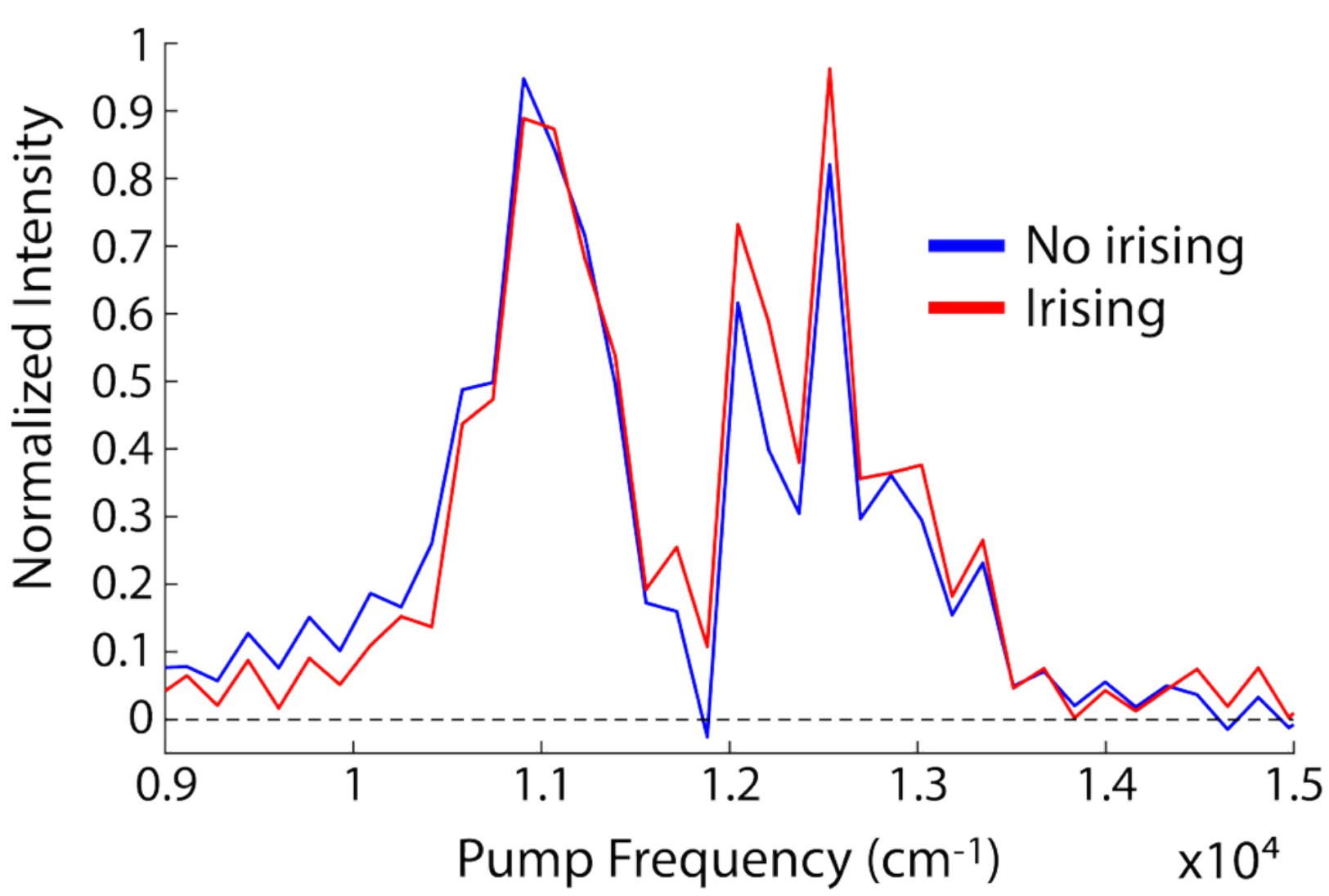

**Figure S1.** Integrated interferometric autocorrelation measurement ("Dark Arm" spectrum) taken of the pump beam before and after irising by a factor of 2-3 before the beam enters the Mach-Zehnder interferometer. Spatial chirp across the beam profile will cause interference, which reduces or completely removes the power in the spectral hole region near 12,000 $cm^{-1}$. Irising the beam mitigates this interference so that the full spectrum has spectral intensity for 2D spectroscopy (see **Figure 3A,C** in the Main Text). A narrower pump spectrum was used for this control measurement to demonstrate the effect of irising the beam profile.

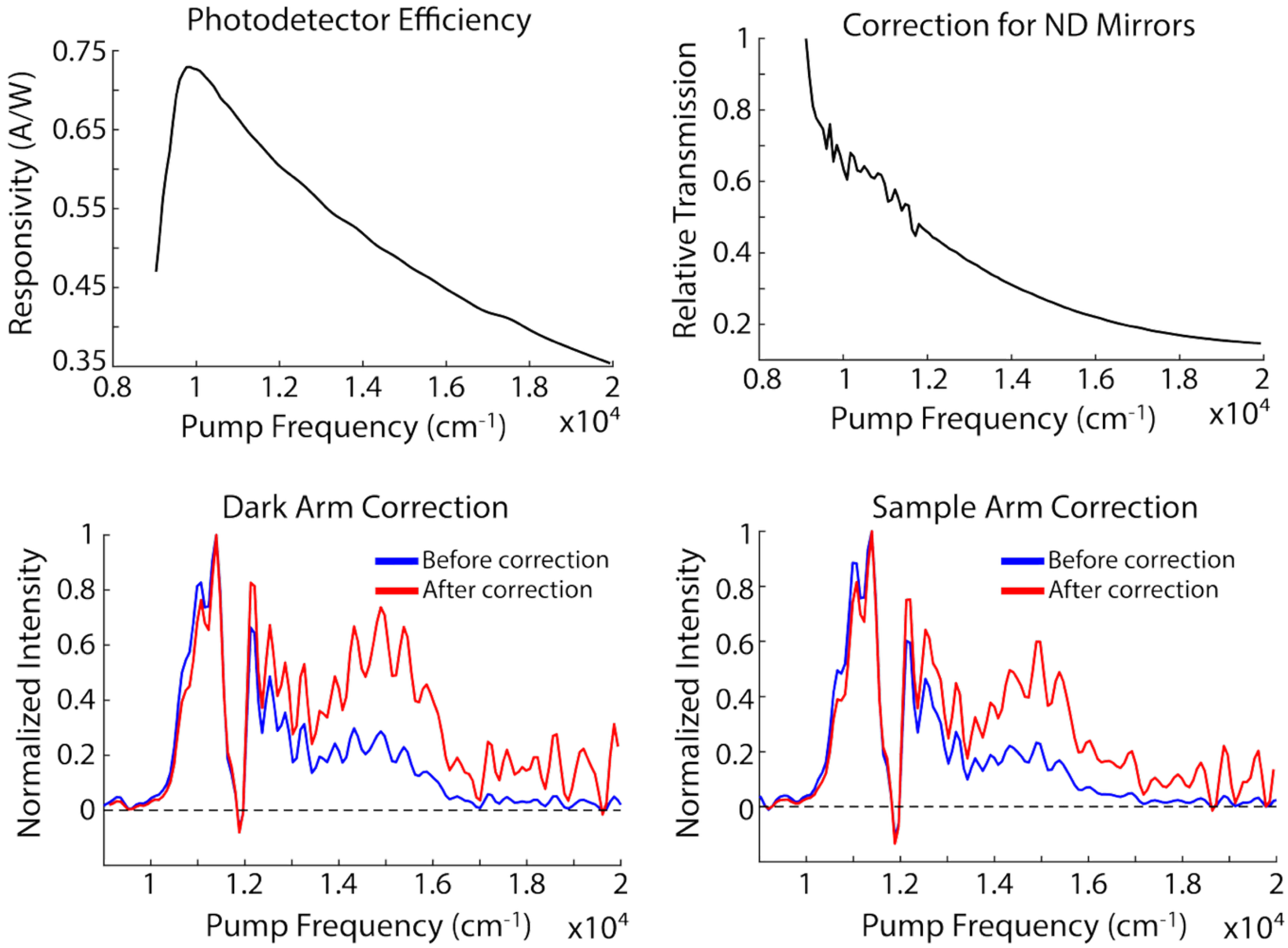

**Figure S2.** Top: Frequency-dependent efficiency curves for the photodetector and reflective ND mirrors used in this experiment. These curves are taken from their respective product details (Left: Thorlabs PDA20X2; Right: Thorlabs ND10B, ND20B, and ND30B). Bottom: "Dark Arm" and "Sample Arm" spectra before and after correction of spectral loss. The correction is made in the frequency domain: After Fourier transformation of the autocorrelation measurement made in the time domain, the frequency-dependent spectral efficiency curves are simply divided from the recovered power spectra. The resulting spectra are normalized to their highest value to illustrate the relative spectral change after correction. The corrected spectra match pump spectra measured using a grating spectrometer (see **Figure 3B** in the Main Text).

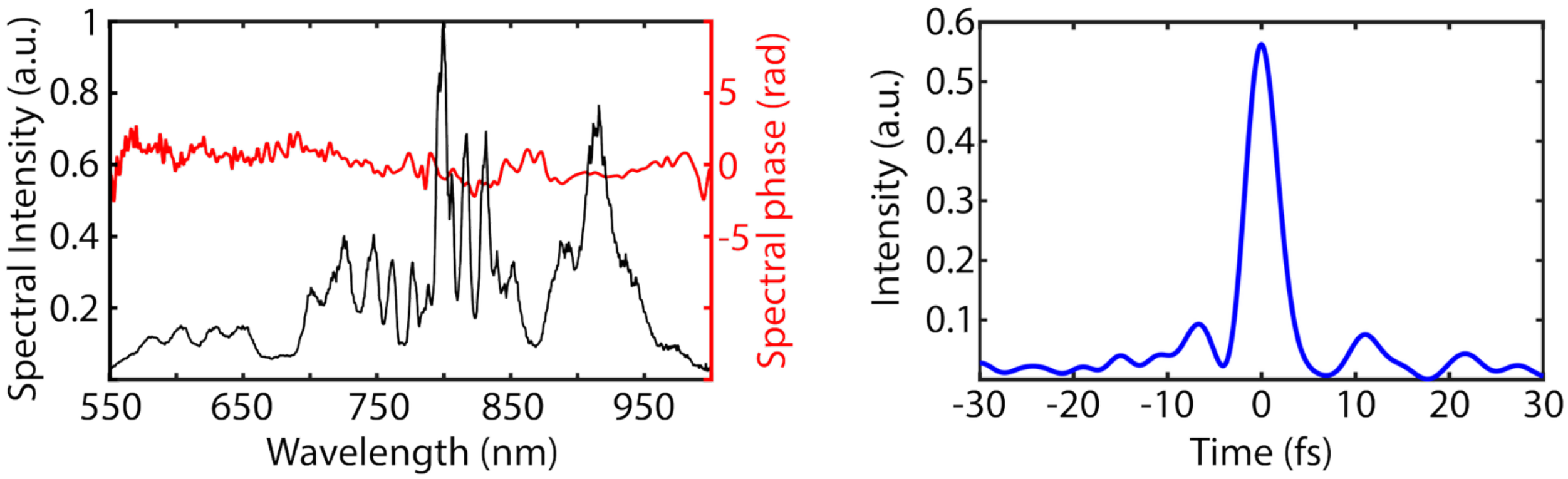

**Figure S3.** Example few-cycle pulse generated from the hollow core fiber pulse broadening setup. The wavelength-dependent phase (left) and time profile (right) of the pulse as measured and retrieved using a D-cycle (see *Methods* in the Main Text). The retrieved pulse width is 4.0 fs FWHM. Note: this few-cycle pulse is not identical to the spectrum used for the 2D EV measurements in the Main Text.

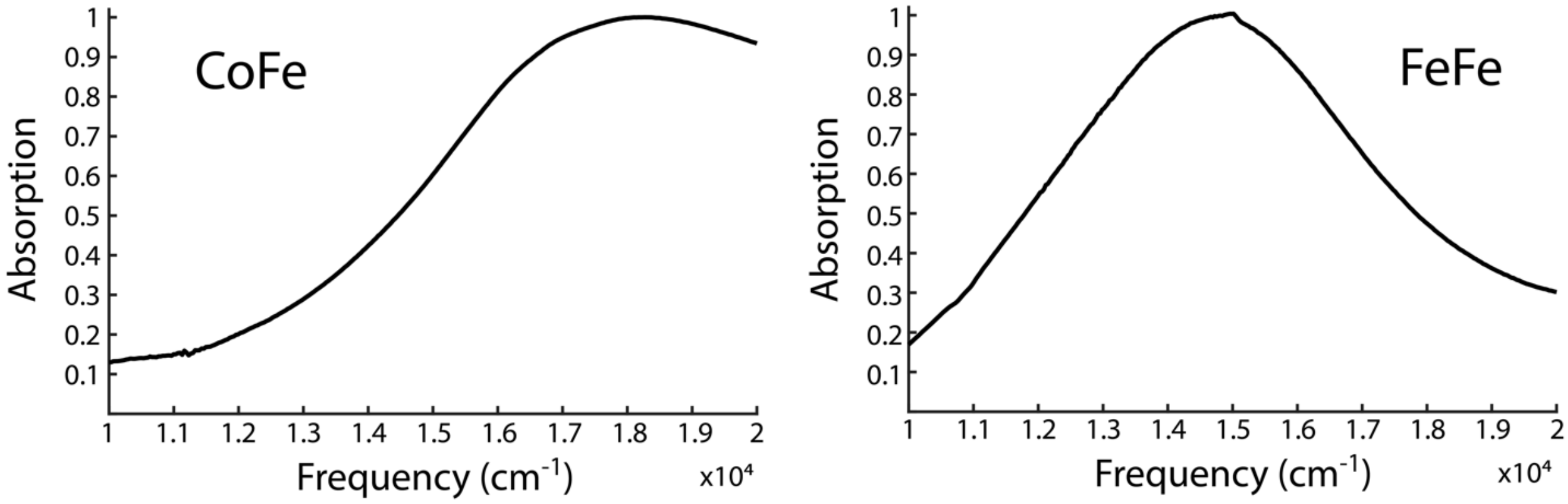

**Figure S4.** Normalized UV-Vis linear absorption spectra of Prussian Blue analog nanoparticles, CoFe and FeFe. The CoFe spectrum peaks at ~18,200 $cm^{-1}$. The FeFe absorption peak is red-shifted relative to CoFe at ~14,300 $cm^{-1}$.